\documentclass[sigchi]{acmart}

\begin{document}

\settopmatter{printacmref=false}
\setcopyright{none}
\acmConference{}{}{}
\acmYear{}
\acmDOI{}
\acmISBN{}
\acmPrice{}
\acmSubmissionID{}
\renewcommand\footnotetextcopyrightpermission[1]{}
\pagestyle{plain}
\title{Choosing Between an LLM versus Search for Learning: A HigherEd Student Perspective}


\author{Rahul R. Divekar}
\author{Sophia Guerra}
\author{Lisette Gonzalez}
\author{Natasha Boos}
\affiliation{
    \institution{Bentley University} 
    \country{Waltham MA USA}}
\email{rdivekar@bentley.edu}
\email{sguerra@falcon.bentley.edu}
\email{lgonzalez@falcon.bentley.edu}
\email{nboos@falcon.bentley.edu}

\begin{abstract}
Large language models (LLMs) are rapidly changing learning processes, as they are readily available to students and quickly complete or augment several learning-related activities with non-trivial performance. Such major shifts in learning dynamic have previously occurred when search engines and Wikipedia were introduced, and they augmented or traditional information consumption sources such as libraries and books for university students. We investigate the possibility of the next shift: the use of LLMs to find and digest information in the context of learning and how they relate to existing technologies such as the search engine. We conducted a study where students were asked to learn new topics using a search engine and an LLM in a within-subjects counterbalanced design. We used that study as a contextual grounding for a post-experience follow-up interview where we elicited student reflections, preferences, pain points, and general outlook of an LLM (ChatGPT) over a search engine (Google).

\end{abstract}





\maketitle

\section{Introduction and Background}

Large Language Models (LLMs) have gained popularity within the student community in the past few years. Many LLMs are freely and readily available to students only using a web browser. Their availability and performance present a possible shift towards using new technology for learning. Specifically, students have been found to use such LLMs for learning complex topics, writing, brainstorming, etc. \cite{divekar2024usage, Albadarin2024}. 

While students are largely adopting AI, the educator community remains divided in their perspectives, ranging from acceptance and the rethinking of educational practices to rejection of the technology and concerns over its use \cite{Crcek2023}. Some educators recognize the importance of preparing a future generation that can effectively leverage artificial intelligence, while others express concerns about academic dishonesty, reduced critical thinking, and the unquestionable acceptance of AI outputs by students \cite{Crcek2023, Kiryakova2023}. Additionally, negative emotions such as anger, fear, disgust, and sadness have been reported, often linked to issues like plagiarism, job security, bias, and disruption \cite{Mamo2024}.

Further highlighting the complexity of AI's impact on education, \citet{Firat2023} found that themes of evolving learning systems and the changing roles of educators frequently appear in conversations with members of the educational community. The notion of systemic change due to AI in education is further supported by a recent survey of U.S. provosts conducted by InsideHigherEd, which revealed that 83\% of respondents reported their universities had published or were drafting policies of AI use in teaching and research, while expressing moderate concern about the risks to academic integrity posed by generative AI \cite{quinn-2024}.On the other hand, 92\% of provosts indicated that faculty and staff of institutions have requested additional AI training, reflecting a growing eagerness to engage with the technology. Further, many professors have already incorporated guidelines on AI-generated content into their syllabi \cite{mcdonald2024generative}, and some universities have provided institution-wide free access to LLMs like GPT-4 and custom computational environments to run them \cite{duraisamy-2023}. Faculty members have also used AI in their teaching and work, e.g., \citet{Mahapatra2024} have used ChatGPT to generate feedback in English as Second Language (ESL) writing classes.

Overall, the education community recognizes that this is not the first time a technology has significantly impacted education, as seen previously with the emergence of Google and Wikipedia \cite{Duha2023}. However, the introduction of new technologies like ChatGPT and their potential at-scale adoption by students prompts investigations of whether they are being used, how they are being used, how tools fit into the learning experience of a university student, whether they replace or augment existing learning processes, and whether they facilitate or hinder their learning process. The outcomes of such investigations can inform how educators can adapt to students using new technologies, reshape expectations for learning, and inform policies that support or limit the use of AI on campuses. Some of these questions have been addressed in the literature. In recent surveys, \citet{Sallam2024}  saw that 85\% respondents had used ChatGPT, higher than previously reported studies. Similarly,  \citet{Chan2023} surveyed students in Hong Kong and found a comparable familiarity, positive attitude toward, and willingness to use generative AI within the student community. Beyond direct learning applications, students have also reported using technologies like ChatGPT for entertainment, mental health support on campuses \cite{divekar2024usage}, and to bridge social connection gaps \cite{Crawford2024}, among others.

Despite these findings, it is not clear how students think about new technologies like ChatGPT in relation to the most popular and well-established use of search engines (e.g., Google, Bing, etc.) for learning \cite{divekar2024usage}. Understanding this relationship can reveal whether there is a shift from current traditional search-based learning strategies toward using LLMs. While previous research has broadly explored students' reactions toward LLMs, our study seeks to understand students' responses when learning new topics–a common academic exercise–using both LLMs and traditional search engines. Given the potential of LLMs disrupting the status quo learning process with search engines \cite{shah2024envisioning}, we directly compare the two tools by conducting interviews with 20 university students to investigate the overall learning experience (e.g., finding, synthesizing, and producing output related to a new topic) with each tool.

Rather than conducting a broad, uncontrolled inquiry, we structure our interviews in the context of a within-subjects counterbalanced study, where students are asked to learn about new topics of moderate complexity using either ChatGPT or Google. Post study, we gathered their reactions across the tools' various dimensions, such as interface, comfort, confidence, trust, ownership, plagiarism, and other learning-related outcomes achieved using these tools. The qualitative analysis of the interview responses and findings is presented here.

\section{Methodology} 
We want to investigate holistically how people feel, use, and integrate learning with LLMs and search engines. However, both LLMs and search engines are versatile tools and can be used in many ways. In addition, learning itself is a multifaceted term. Therefore, we needed a specific context in which students could be interviewed. We used an experiment-style counterbalanced protocol to create that context. In the protocol, participants experienced each tool in a learning context. They were then asked follow-up questions to reflect on their experience and connect it to the larger use of the two tools in their broader student life. 
 
Concretely, the study was designed as follows. An initial intake survey was offered electronically to undergraduate and graduate mailing lists at a doctoral-granting business school in the northeastern U.S. Students who were currently taking courses or indicated that they plan to take courses with the PI were excluded in the interest of potential harm to students and the validity of the results. Of the several respondents, 20 were chosen based on their completeness of responses and higher self-rated expertise with AI tools for learning. We scheduled 1:1 sessions with 20 participants. In each session, after completing initial steps such as greetings, consent forms, explanation of procedures, etc., participants cycled through 7 phases of the study.  In phase 1, a participant was assigned a topic and given up to 10 minutes to write about it without any tool other than a word processor to determine their existing knowledge. Topic assignment was a randomized choice between ``How does the internet work?'' and ``How does the electricity grid work?'' as we assumed that they are similar in difficulty and familiarity for our population at a business school.  This phase allowed us to confirm that participants' existing knowledge of the topic was negligible.  In phase 2, they were provided another time box of 10 minutes to learn about the topic given a pre-assigned tool (counterbalanced assignment between ChatGPT and Google). For example, a student was asked to learn how the internet works by using only ChatGPT.  In this phase, they were allowed to take notes using a word processor. In phase 3, they were asked to write about the topic again without any tools to determine their post-learning knowledge; they were allowed to consult their notes. Then, they cycled through phases 4-6, exactly the same as phases 1-3, except that the tool and topic were switched to the alternative not initially picked. 

Learning phases 2-6 provide an approximation of the learning process consistent with the common Input-Processing-Output framework for learning, where a student has to find and expose themselves to new knowledge (input), critically analyze and connect information (process), and then produce/demonstrate their understanding (output). This provided a learning context for probing the questions in a semi-structured interview in Phase 7. Our interview questions were designed to elicit their reflections on their experience, whether they felt ownership over the work and process, and their confidence and satisfaction using either tool. The questions also elicited responses beyond the study and in general, such as how their learning process has changed after ChatGPT was introduced, where they learn about this technology, how they perceive the technology, and their own expertise using it. The complete interview guide is attached as supplementary material. The participants were compensated with a USD 50 gift card. The university IRB approved the project and appropriate consent was obtained from the participants.

We automatically transcribed and manually coded the interviews. Four people were involved in the coding process. Initial codes were developed by the PI based on the questions in the interview guide and familiarity with the collected data. The codes were primarily developed to organize responses to themes of questions. 
The meaning of codes, along with one coded interview, was explained to the rest of the team. A few more codes were added, resulting in a total of 131 codes, including sub-codes; thereafter, no more codes were necessary. Rather than a standardized coding process, we aimed to maximize coders' diverse perspectives and interpretations. Therefore, multiple coding was encouraged, and agreeing on codes was not a priority. As a result of our methodology, rather than reporting trends, we report every theme that adds to the body of literature, even if it was only one participant reporting it. We conducted coding in NVivo to group responses by questions. Then, we found themes of responses to each question, grouped them for readability, and present them below.

\section{Demographics}
All participants were enrolled in [Anonymized] University in the northeastern region of the United States where the study was conducted. Table \ref{tab:demographics} shows the self-reported demographic information of our sample. As we see, most participants were between 18-20 years old 11 female and 9 male undergraduate students. Most students were in good academic standing, as seen in their self-reported GPAs, and indicated business-related topics as their major, minor, or area of expertise. This context is important as findings for another demographic may be different, e.g., \citet{zamfirescu2023johnny} have found that technical students might be using AI tools differently than nontechnical users.

\begin{table}[!htbp] 
\centering
\begin{tabular}{l c}
\hline
\rule{0pt}{10pt}\textbf{Description} & \textbf{Number of Students} \\ \hline
\\

Age Range & \\ 

\hspace{2em} 18-20 & 10   \\ 
\hspace{2em} 21-23 & 7 \\
\hspace{2em} 24-26 & 3 \\

Enrolled Level of Study & \\
\hspace{2em} Undergraduate & 13 \\
\hspace{2em} Graduate & 7 \\

Gender & \\
\hspace{2em} Female & 11 \\
\hspace{2em} Male & 9 \\

GPA (max. 4) & \\
\hspace{2em} Above 3.5 & 14 \\
\hspace{2em} 3-3.5 & 5 \\
\hspace{2em} 2.5-3 & 1 \\

Majors, Minors, or Areas of Expertise & \\
\hspace{2em} Finance & 6 \\
\hspace{2em} Management & 5 \\
\hspace{2em} Business Analytics & 4 \\
\hspace{2em} Marketing & 3 \\
\hspace{2em} Accounting & 2 \\
\hspace{2em} Corporate Finance and Accounting  & 1 \\
\hspace{2em} Computer Information Systems & 1 \\
\hspace{2em} Economics & 1 \\
\hspace{2em} Health Studies & 1 \\
\hspace{2em} Info. Design and Corporate Comm. & 1 \\
\hspace{2em} International Affairs & 1 \\
\hspace{2em} Sports Business Management & 1 \\

\hline

\end{tabular}

\caption{Participant Demographics}
\label{tab:demographics}
\end{table}

\section{Qualitative Analysis}

In this section, we describe themes found in interview data from participants. Participants are labeled as \textit{P} followed by a number; e.g., P1 would mean participant one. Participant quotes exemplify the themes but do not indicate their popularity unless specified. Some quotes have been lightly editorialized to correct the ASR output and remove filler/repetitive words.

\subsection{Comparing Interfaces, Inputs, and Outputs}
We found that the students had mixed responses in their outlook towards ChatGPT and Google. Here, we describe various reasons (themes) related to the interface, input, and output that contribute to a positive or negative experience with either of the tools.

\subsubsection{Interface}

Participants liked ChatGPT's conversational interface. P11 said, ``It's like you're speaking to a person.'' Further, P11 mentioned, ``[ChatGPT is] like, `yes, I can help you with that,' and it makes you feel more comfortable, and it helps,'' acknowledging the wrapper of polite responses produced by ChatGPT that enhances the conversational user interface.

P1 said they liked ChatGPT because it provided a continuous interface to the information-consuming experience, ``It is easier to find notes like these because it is just one web page where you are typing and it's answering and then you're asking more questions and creating your notes.'' 
P22 said, ``ChatGPT is quicker, easier to use, just like easier interface. It's more easy to read. They lay out for you in bullet points ... the design of like the ChatGPT website is easy. ... It gives you that one answer that you need.''

Participants also commented on the features that improved usability, such as permanence of output and accessing multiple chats. For example, P22 added for ChatGPT ``something that's like aesthetically pleasing and easy to like, use, and stuff like that. Yeah, I would say I like how you can save your chat, your chat boxes in ChatGPT, like I have different tabs now. It's like your work doesn't disappear.''


\subsubsection{Relevance of Output} Participants seemed unsatisfied with the results of Google search, where the top results may not give them exactly what they want. P9 said, ``With Google, you just can't get very specific information that you want, and you kind of have to piece together certain information.''


P22 mentioned they liked that the ChatGPT output was localized to them and that ChatGPT assumed some of the context, ``I was going to look up [the frequency of electricity] in ChatGPT, but ChatGPT already said ... for the U.S., it's 60 Hertz. So ... it just recognized that we were in the United States.''


\subsubsection{Scaffolded output:} Participants mentioned that Google often leads to too much information. For example, P22 said, ``Google, it's like overwhelming. There's just so much to look at when you ask one simple question.'' 

In addition, the information is not necessarily adapted to one's reading or understanding level. P11 noted that ``[Googled articles] have a more difficult language, more professional formal language. People that publish articles they are highly knowledgeable, and they have a a great vocabulary, but sometimes it's not the best for you as a reader, especially if you're a new guy in a certain [topic] area.'' That changes with ChatGPT as P23 pointed out that they liked the simple explanations and an option to make it simpler if required: ``The simplest words explain to me what the power grid is ... like, [I can say] `explain this to me in simple words' or `how you would explain it to like an eighth grader'.''

\subsubsection{Variety in output:} P21 pointed out that the variety of information provided is significantly less with ChatGPT. ``The big downside [with ChatGPT] is that I don't have many options to besides the one single like generated answer.'' ``Whereas with Google ... these are such big search engines as kind of like you can just put in one word and it will throw in so many, many things for you.''(P3) 

However, most of the people had a different opinion that the variety of output can become overwhelming. P22 said, ``Google was all over the place.'' P23 frustratingly mentioned, ``I felt like it did not give me exactly what I wanted. I want to know how the Internet works and it gave me like going through three different websites. It gave me so many different, different, different explanations. I was like, OK, what's the network like? What are packets? Just tell me like one thing, right? Give me one process. I want to understand it and I want to like then go ahead and research more about it. But for the basics, I want to just understand it in a few simple words.''

\subsubsection{Type of input: } Participants mentioned that with Google, they have to enter keywords rather than full questions on Google. For example, P10 said, ``I can't really look up a question [I have] to more so look up a subject and then read like a whole article or website to try to find the answer.''


\subsection{Comfort, Confidence, Satisfaction, and Trust with tools} 
Here, we explore responses to questions that elicited participants' comfort, confidence, satisfaction, and trust with the two tools. 

\subsubsection{More confident using ChatGPT for learning because it makes sense} Participants felt more confident in their downstream output-related work (e.g., presenting/writing on a topic) when they used ChatGPT because ChatGPT's output was more understandable. P11 said, ``If you're reading a paragraph and it doesn't make sense ... and if the professor asked you a question about that ... I could try to explain it, but [I] still lack confidence ... When you have consulted ChatGPT, and you understand the basics better, you are able to sound more confident when you're actually providing your professor with an answer or when you're doing anything.''

\subsubsection{Less confident with ChatGPT because of accuracy  issues} Participants acknowledge that they feel less confident using ChatGPT while looking for accurate information, e.g., P9: ``If I want like actual facts, I never really use ChatGPT because ... some of the facts just aren't right'' and especially in high-stakes scenarios, P22 said, ``If I'm going to submit something, I want it to be as accurate as possible because I don't want my professors thinking  like `what is this?'''

\subsubsection{Less confident with Google because self-doubt in selection of information sources} Some participants acknowledged that search results can also be wrong and picking the right from wrong information adds cognitive load, e.g., P22 said, ``So it's more just like, do I trust myself and what I'm reading and picking ... just because there's so many results that [Google] spews out. So I'm like, am I picking the right thing? ... Is there a better place I could be looking for this?''

\subsubsection{Confidence in Google affected by ads} P14 mentioned that for Google, distrust forms because of ads, ``the first two things that it shows you are normally ads. So that kind of makes me distrust it more because often those are what you think you're looking for, but you often have to scroll down more.''

\subsubsection{More confident with Google because control over sources} Participants considered Google as more reliable as it offers more variety and control over information sources than ChatGPT. P10 mentioned, ``I'd argue Google's a little bit more reliable, especially 'cause you can assess where the information is coming from'' whereas, ``ChatGPT will give you information and you don't really know what sources it comes from. If you ask for the sources, it, it still won't really give you like direct links.'' P3 mentioned ``I would personally prefer Google because there's such a variety.''

When using Google, participants applied various strategies to select sources and took advantage of information source control, e.g., P18 mentioned, ``I check the first link first'' because it was perceived as ``what Google recommended.'' Some select based on familiarity with the brand; P11 said, ``One of the websites that showed up was hp.com, and obviously HP is a well-known company,'' or based on perceived provenance, e.g., P14 mentioned ``I used two .gov websites. I tried to avoid information which I knew could be changed. So I didn't go on to Wikipedia because I know that could be changed'', or based on the automatic summary pane of Google (pre-Gemini), e.g., P16 said, ``[Google] gives like the little blurb from the article. I read the blurb and I'm like, that seems like what I need. So I just clicked on that article and it gave me what I needed.''

Participants also intentionally avoided specific sources, e.g., P22 said, ``I don't really take my information from the blog or maybe from like news sources like New York Times. I'll use like Google Scholar.'' Yet some are unsure if the source is good enough. ``Although I have my own criteria, like is there a better place I could be looking for this?'' and can find the number of sources to be overwhelming, e.g., P21 said, ``I had many sources that they don't know which one to pick from.''

Regardless of the strategy applied, the agency to select was appreciated. As P11 said, ``You still have to be careful on what websites you're choosing to refer to so, but yeah, it's up to you though. It's not up to the tool. With ChatGPT, you don't get a chance.''

Many participants noted that Google's transparency of sources brings trust. As P21 said, ``It all comes down to the sources as well. I think sources that I already knew were trustworthy just to save myself that like extra time of thinking if this is right or wrong'' and with ChatGPT, P18 said, ``I would have wanted to turn to Google to verify everything.''
P18 also said they would rather trust [an] anonymous human than a machine: ``Weirdly enough, I'm more inclined to trust a random user on a random forum than I am to trust ChatGPT.''

However, with Google, participants noted that not all links can be accessed, e.g., P5 said, ``So with Google, first of all, a lot of these articles aren't free.''

\subsubsection{Confidence affected by tool  familiarity}
Participants mentioned that having used Google for many years has taught them to sort through information, and that gives them confidence. P14 said, ``I think just because, again, I've used Google my whole life, I feel confident finding stuff on Google.'' P5 said, ``With Google, I think the information is there. You just have to know which one is the right information. And because we've been using Google for so long, I think it's easier to sort of, yeah, figure that out.''

\subsubsection{Domain knowledge affects trust and satisfaction with ChatGPT} 
Several participants mentioned that their prior knowledge of the topic they are looking up on ChatGPT affects their trust. P9: ``I think [ChatGPT is] trusting when ... you kind of have like background knowledge on [the topic]. It is not as trusting when you're relying only on ChatGPT  for all the information you're getting because you can never tell if it's wrong or not.'' In addition, P22 mentioned not knowing enough can lead to a cold-start problem, ``I feel like once you figure out the right questions to ask, it can be really helpful. But how are you supposed to know those right questions, right?''



\subsubsection{Confidence affected by lack of transparency in thought process and discussion} P18 mentions that ``Forums can be really good for me understanding something because again, people explaining it to others tend to make it easier to understand. So I think forums can be really useful for that, but you can't get that on ChatGPT where it'll just confidently give you an answer regardless of correctness.''

\subsubsection{Not matching user's voice leads to lower satisfaction} P16: ``I prompted it to give me like an outline for an essay. And I feel like the three points it gave me were like amazing points [but] like it didn't highlight the points that I wanted to talk about. So I felt like it kind of, it doesn't understand the importance of some things.''

\subsection{Synthesizing information}

ChatGPT's ability to synthesize information was the most common positive.  P22 said, ``I use ChatGPT in an academic setting because it's so good at synthesizing information, and my brain doesn't do that.'' We discuss specific synthesis features that participants called out.


\subsubsection{Prioritization of information} P21 mentioned they liked ChatGPT's prioritization of information presented: ``I think especially for the first [topic I learned], ChatGPT helped me a lot [with] understanding what was the most important part by asking repeatedly,  what was the most important things ... And it made connections for me to like not really have to do it myself.''

\subsubsection{Need for a single narrative answer} P1 said, ``If it is a professional problem in terms of career, [or] academics, office work, everything, then I would always prefer to go to ChatGPT  because I know that I'll get a definite answer.''

\subsubsection{Better than self-synthesizing visual input} P21 said, ''I had like an issue when [using Google] I was like, I got an infographic from the website and then adjusting that to like actually write with like one document just didn't work. So I just like worked through that, which if it was like ChatGPT, I would have just copied and pasted like a line of text.''

\subsubsection{Low risk, high reward synthesis} P10 mentions that even when search/synthesis goes wrong with ChatGPT, the invested effort is worth it. ``I feel as if I can get a lot more information a lot quicker. And ... even if the information isn't what I'm looking for, that time spent, or I guess the lack of time spent for it, makes up for it.''



\subsection{Prompting Knowledge Affects Experience}

\subsubsection{Prompting to get deeper}
Participants noted that knowledge of prompting can affect results, especially while getting deeper into the topic. E.g., P3 said, ``So I feel like the prompts do have a huge impact on like the overall experience because if you really have a substantial amount of knowledge on something, AI doesn't really help you with [going further without prompts].''

\subsubsection{Uncertainty in prompt formulation} Participants express uncertainty about whether they are asking the right questions or formulating prompts correctly to get the desired information. This is evident in quotes like P23 said, ``I just feel like, OK, I’m putting the prompt in, but are there more meanings to what I’m asking?'' and ``I’m not sure if I am, you know, correctly prompt engineering.''

\subsubsection{Time and effort in crafting prompts} Participants mention that it can take extra time and effort to formulate the right prompt to get the desired information. This is illustrated by P3, who said, ``Sometimes it takes me extra long to think of a question or like to type in the correct question'' and P21 said, ``I had to, like, try ten different things and [ChatGPT] just never gave me the answers.''

\subsubsection{Iterative prompting process} There’s a recognition that prompting ChatGPT requires an iterative approach, asking follow-up questions or reformulating prompts to get more comprehensive or accurate information. This is seen in statements like P20, which said, “And then the ChatGPT had to like figure out, OK, I’ll ask one question and then ask a second question to see if like give me a different set of information.”

\subsubsection{Comparison with Google’s ease of use} Users often compare the prompting process in ChatGPT with the relative ease of using Google, where less specific queries can still yield useful results. This is evident in quotes like  P19 said, “You don’t have to spend a lot of time like really like altering the way that you ask the question to Google.”

\subsection{Goals and Task Expectations affect Experience}

The goals and expectations of an academic exercise affect whether students will choose ChatGPT or Google. We explore various themes related to it here.

\subsubsection{Type of topic researched} Participants recognize that well-established knowledge is better looked up on ChatGPT as P16 says,  ``Like just for today's work, I feel like the information's been out there on the Internet for so long. Like, I don't think there's any way I could have pulled this false information on how the Internet works or like power grids work 'cause that's just, it's such a known topic at this point.''

\subsubsection{Depth of information desired}
Participants mention that if depth and variety are required in research, ChatGPT may not be the best tool, as P23 says, ``You want to read different kind of articles, get different opinions ... When you're going in depth sort of researching, that's when you want to use Google and get those opinions and views from different websites.''


\subsubsection{Goal of learning versus producing output} Participants recognize that learning and producing output can sometimes be different and that the tool is chosen based on the goal. As P18 says, ``I think for this [study] specifically, I liked Google better because for this it was more of a learning rather than necessarily a writing exercise.''

\subsection{Learning with ChatGPT and Google}
Participants generally found being ``more productive using ChatGPT'' (P22). We explore their reasons here.

\subsubsection{ChatGPT provides a structure} Similar to synthesizing information, participants mentioned that ChatGPT helps them prioritize subtopics and feels more structured. P18: ``Much more structure than Google and help me kind of identify three big parts of the power grid generation, transmission and then the distribution.''


\subsubsection{Learned more details with Google} In contrast, the students liked that Google offered more detail and depth of learning. P9: ``But then the information that I got from Google definitely went way more into detail and like actually explained the process and ChatGPT was very vague with it.''; especially without much user-initiated follow-up ``like when I was doing the power grid and [ChatGPT] talked about, Oh, like it sends energy and it connects the producers and consumers. And like usually Google site would probably explain what that is. But then I have to ask you like ChatGPT again.''

\subsubsection{ChatGPT provides writing and articulation support} In addition to learning, the participants mentioned ChatGPT offering significant help with writing that was sometimes better than their own; P10 said: ``I feel like I don't even usually write that well on my own.'' Another participant said that long-term use has affected their articulation abilities too; P8: ``I guess it's just evolved the way I speak and the way I articulate myself too.''

\subsubsection{ChatGPT provides a contextual follow-up}
Participants mentioned that they liked the ability to ask specific follow-up questions that allowed them to deep explore the topics. For, P1 said, ``It will give you everything in one go, and then you can actually build on that conversation and understand and have a deeper understanding ... So I feel they're helping me.''

\subsubsection{More incidental learning with Google}
Here, we define incidental learning as unplanned exposure/learning of new topics rather than the broader definition of learning while not expecting to learn. The participants mentioned that they incidentally learned about more topics on Google than on ChatGPT. P11 said, ``Yeah, even right now when I was doing the research [using Google] for Internet, how it works, I was jumping from one article to another and one website to another, and sometimes you just come across something that you were not even looking for, whereas with ChatGPT I guess if you require information about a certain topic it's going to stay [at that topic]... in Google you can just `oh that looks interesting, let me click that`. ''

P22, however, mentioned that not all incidental learning on Google is productive, ``it was going into like some like weird sections of the Internet because I'm asking such a bad question of like how the Internet works. That's another thing too, where it's like there might be some conspiracy theories on Google and I'm like, I don't want this.''



\subsubsection{ChatGPT provides learning with on-demand contextual examples} Participants mentioned that in addition to asking questions, ChatGPT is also great at generating examples that help digest complex topics, e.g., P23 said, ``And it's giving me  the example that I could easily compare to and understand within a few seconds.''

\subsubsection{Questionable Retention with ChatGPT} However, some participants questioned how much of their learning they will retain. For example, P21 said, ``Honestly, I felt pretty good given I had like a page and a half of notes from the topic I didn't even know at all 10 minutes before. So that was a little bit good. But I don't know, I would definitely like to go back and just see how much of that is actually important and how much I actually like understood or I can  retain afterwards too.''

P6 sheds more light on retention and associates it with the effort they had to put in to synthesize the information ``So instead of really like absorbing the information, I was more just kind of getting it to write it for me and then do it ... I'm probably going to remember more about the power grids than I am about the Internet because I took the time to like, look at different sources for that.''

\subsubsection{Harms learning ability} 
Several participants perceive that ChatGPT harms their learning ability using words like lazy, brain rust, etc. E.g., P15: ``So definitely it's a good tool, but it does make us lazy in terms of like finding our own information and learning.'' P13 even says that they don't learn because it is an easy way out of putting in the work ``I just don't learn anymore ... if it's really tedious [work] and I don't feel like I'm going to take a take away from it, I'll just use ChatGPT, get it done and get the grade.'' P9 expands more by pointing towards ChatGPT circumventing the learning process ``Like I think [summarizing information is] a really important step in link researching and processing information. So then I think it really like limits my overall learning when I'm trying to get away with the easiest option.''



\subsubsection{Surface level learning with ChatGPT} Participants mention that if they learn, it is at the surface level. For P9 said, ``I'm not getting an in-depth learning. I'm just trying to do in a more overview and like surface level, and it's way less in-depth now.''


\subsubsection{Tools are similar but not replaceable:} Several participants acknowledged that although ChatGPT and Google can be used for similar purposes, they are more suited towards certain tasks. For example,  P19 said, ``I know that you can use chat[GPT] kind of for almost the same purposes as Google. But to me, I don't think it's like replaceable ... in the end, you still have to kind of put in the work to make sure the information is correct. You still have to like reword stuff to make sure it's not plagiarized. Like the way that ChatGPT  words things  very much sounds automated. And like the words that it uses are very like complex to the point where you kind of have to sit like manipulate it on your own.'' Whereas on Google ``you can't type in like a paragraph and say like, give me like a topic, like a title for this or like reword this sentence for me.''

P18 said they use ChatGPT to augment their Google search while learning ``ChatGPT was really good at giving me keywords that I wouldn't think to search about. Google was really good at giving me information when I had a keyword and then delving deeper into that information.''

\subsubsection{Using other sources to Verify ChatGPT} Participants mentioned that they use Google (and other tools) to verify the output of ChatGPT when it matters, e.g., P23 said, ``So ChatGPT basically breaks down the concept for me but understanding if that breaking down is correct is something I get from Google and more so I would say Youtube.''


\subsection{Other uses of ChatGPT beyond the study}
We asked students how they use ChatGPT in their daily life for education and university-related purposes. Here, we explore the themes of their responses.
\subsubsection{ChatGPT available as a quiz partner} P22 said they use ChatGPT to deepen their understanding by asking it to quiz them - ``I have asked ChatGPT to create like mock quizzes for me to help me study. I always tell them like just my roommate or my friend, like, hey, like just I have my study guide ask me questions.''

\subsubsection{Brainstorming} P11 mentioned that they used ChatGPT to think more creatively by brainstorming. ``You can just use ChatGPT to do the brainstorming, or it could also help you extend [the] brainstorming phase. Because ChatGPT is going to come up with a list of 10 things  that you might or might not have come up with, and in those ten things, there are things that lead you to other things as well. They're like, OK, marketing. How can I relate this to that? And it actually sends a list to like 20 things. And then you just narrow it down with whatever you need to do and you just pick two or three that you need to work on and then move on it like accelerates the process.''

\subsubsection{Document synthesis assistant} Participants mentioned they use it to ask questions or summarize a document or web articles. P21: ``I always have like chat on this side, even for just like reading articles with like I could just like read an article to copy paste it and be like, what do you think's most important? And then see like how my notes compared to chats and then having a good balance there.''

\subsubsection{ChatGPT available as a Personal tutor} For some, ChatGPT acts as a personal tutor for math and coding tasks. E.g., P14 said, ``If I wanted help in a math problem, instead of going in [to a teacher], I could ask ChatGPT. It could give me like how I do it step by step ... So I think it has changed how I learn as I have like practically a teacher on my computer that I can use to help me complete homework.'' 
P1 mentions that it helps augment office hours; ``Sometimes the professor has already given a support and I'm not understanding why this particular [code] function is used. So then GPT will tell me, you know what, this is used because of this and etc etc.'' P24 mentions it helps them clarify concepts from lectures, ``some concepts which I might not understand in class, and I might just ChatGPT them ... it might just give out something which my brain understands much better''

\subsection{Plagiarism Anxieties} Several participants raised concerns concerning plagiarism and have different opinions on whether ChatGPT-generated output can be caught.

\subsubsection{ChatGPT is easier because plagiarism CANNOT be caught}P13: ``I feel like ChatGPT is easier because it doesn't have like  plagiarism. [With] Google when you copy and paste you'd have to actually reword it.''

\subsubsection{ChatGPT is harder because plagiarism CAN be caught} P22 said, ``If I'm getting graded [I'll use] Google. I'm not using ChatGPT. Not because I think my professors have a tool that analyzes my documents and they know because I, I highly believe they don't that they have something that like double checks it and they know it's AI because AI is so new.''

\subsubsection{ChatGPT is hard because citations are unavailable} P19 said, ``if I'm going to write a paper or something like that, like I'd rather not put the risk in using like ChatGPT to give me that information and  get in trouble or something like that versus using like Google and like being able to cite it.''

\subsubsection{Unsure if it is plagiarism:} P20: ``Halfway through [the study],  I started thinking like when writing that I was saying [to myself], `wait, am I supposed to like, like am I allowed to just like write what I copied and paste or am I supposed to like, you know, like try to rewrite what like ChatGPT gave me.' '' P22 mentioned ``It feels like cheating in a way.''
\subsection{Perceived Ownership and Contribution Over Produced Material}
We asked students if they feel ownership and how much is their contribution to the ultimate essay they produced. We found the following themes.

\subsubsection{Ownership with ChatGPT because participant verified information} P18 took it further and said, ``I think part of the ownership now comes from being able to vet the information and use either prior knowledge to know if it's true, use outside research to know if it's true.'' They mention that if ChatGPT was perfect, they wouldn't have felt ownership by saying, ``If I was 100\% confident that it was always correct, I would feel almost no ownership because then it's like the equivalent of copying an article outright and then tweaking some words. So that gives me more ownership because I have more input in that sense.''

\subsubsection{More ownership with Google because participant filtered and paraphrased information} P3 mentions that ``using Google, for example, I had to manually filter [the information]. So although it was ownership from the Internet, because it wasn't my own direct words, I could still filter it in a way to like kind of paraphrase when needed instead of just like kind of copy copying word for word.''

\subsubsection{Ownership with ChatGPT because participant prompted it} P5 mentions that their human agency and initiative to prompt ChatGPT makes them feel some ownership, e.g., ``I feel like I did because I put into prompts ... So I did have some ownership over it.''

\subsubsection{Ownership with ChatGPT because participant synthesized final output} P23 mentioned that ``though the notes are pretty much completely ChatGPT, what I put in the [final] essay was what I understood out of those notes and understood out of what ChatGPT gave me.'' Thereby connecting ownership to their information synthesis effort. 

\subsubsection{Ownership because ChatGPT is not human} P9 disregards ChatGPT as an author; therefore, there is no one else to credit ownership to ``but because it doesn't have like the author or where it's from or what the research is from,  I feel like more ownership.''

\subsection{Sources of AI information}
So far we have discussed the in-technology experiences with ChatGPT and Google. Here, we explore another aspect of ChatGPT adoption in universities: learning about the tool and how to use it.

\subsubsection{Introduction to ChatGPT}
When asked how participants were introduced to ChatGPT, they mentioned the following sources.

\textbf{Co-workers} Several participants got to know about ChatGPT through co-workers at their internships or their prior jobs. P1: ``So I was working on a start-up in India, and I was in the tech team ... that is how I was introduced to GPT,'' and P22 said, ``With the other interns were using it.''

P6 learned about it because of their internship organization's initiative to improve ChatGPT adoption, as they said, ``The marketing team I was working with was trying to keep up like a lot with all the advancements in AI and encouraging to use it ... like ChatGPT or Bard.''

\textbf{Friends, Classmates, and Common Culture} Some participants learned about it directly from friends; as P11 said, ``When I  came here, I was an international student, and I didn't even know to do the I don't know when it was invented ... and everybody was using it every all my friends were using it. ''

P12 hints towards a broader conversation that students are having amongst themselves about ChatGPT, and that is where P12 learned about it. They said, `` It became like a whole thing even amongst my friends, everyone was using ChatGPT.'' with P13 echoing ``Everyone just talks about it.''


\textbf{Social and News Media:} Participants first learned about it through various social media and news platforms like Tiktok, Instagram, etc. P16 said, ``I saw a Tiktok about it when it first came out. And I'm like, oh, I could use that.'' and even informally through memes, as P5 says, ``There are memes about it, people who experiment like trying different prompts.''

\textbf{Professors and School Teachers:} Some participants learned about it via their classes at the university, e.g., P19 ``took an AI marketing class with [university professor's name]'' whereas others learned before they came to the university when they were in school, e.g., P9 learned about it in ``school because teachers started talking about how you're not allowed to use it.''

\textbf{Parents:} P9 also learned about it from their ``dad [as] he works with technology in a school district.''

\subsubsection{How did they learn to use ChatGPT?}
We know that some students mentioned that they need to learn to prompt them.  We asked them where they learned to use ChatGPT and explore the responses here.

Similar to the introduction to ChatGPT, participants learned how to use it from 
\textbf{Friends: } ``[friends] teaching me how to, like, write an essay with it'' (P10) or via \textbf{Self-exploration:} ``I just opened an account and just used it'' (P11) or on \textbf{Social Media:} ``Go on Reddit'' (P18) or in \textbf{University Classes:} ``Took an AI marketing class'' (P19) or using \textbf{Online Information:} ``Found out a lot just like reading about it too. And then I, there was this video that I saw like maybe five months ago about like prompt writing itself'' (P21) or from a \textbf{Parent:} ``And [dad] was like, he's very like, for like using it, but he's like, tries to like explain to me how I feel like use it the right way'' (P9).


\subsection{Privacy Perception of Tools}

\subsubsection{Privacy doom and helplessness}
There is a general helplessness towards privacy that several participants showed, e.g., P24 said, ''To be very honest, I'm concerned about any privacy, whether it be phone, Google, anything ... somebody's tracking you, but what can you do about it?'' Possibly because they feel it is too late to be worried about privacy, e.g., P16 said, ``Either with Google or ChatGPT, every company has my data already, so it does hardly matter to you.'' While others did not give it another thought, e.g., P10 said, ``I haven't really thought about that.'' 




\subsubsection{Low-stakes data}

Some participants mentioned that especially in the context of education, there is not much data that is high stakes, e.g., P21 said, ``Really wouldn't mind people knowing that I've looked up like what is a power grid.''



\subsubsection{I benefit from it}
Some participants said it is okay because they benefit from it, e.g., P18 said, ``It's whatever they use my data, I guess to probably train future models, but cool, make the tool better benefit for me.''

\subsubsection{Security matters more than privacy}
For many students, the security of their data mattered more than the monetization of their data, e.g., P19 said, ``The only thing I'm ever really concerned about is like passwords and like credit card like information.''



\subsubsection{Feeling secure through shared vulnerability} Participants find comfort in knowing they are not the only ones in this privacy doom. As P11 said, ``Everybody's on the same boat because everybody has a Google Gmail account, and so if my data is at risk, everybody's data is at risk. So I don't believe that bad things would happen to everybody in the world.''

\section{Discussion}
Overall, we saw mixed responses towards ChatGPT and Google. Here, we discuss the findings in light of the existing literature and opinion. 

\subsection{Interfaces, Inputs, and Outputs}
An interface and the type of input and output it allows are directly related to usability and user satisfaction. Readers who have used ChatGPT or Google will recognize that, while both are primarily text interfaces for learning, the former allows for a conversational back-and-forth, whereas the latter is a query-response-based interaction paradigm. Several studies have discussed that conversational user interfaces (CUIs) can improve user engagement and satisfaction, reduce cognitive load, and make information more memorable, specifically in a Search as Learning (SAL) paradigm \cite{wu2024promoting, qiu2020conversational}. In our study, we find that higher education students have mostly positive reactions toward ChatGPT, a CUI. Further, they acknowledge that the CUI design that includes linguistic markers exhibited by ChatGPT to signal politeness is well received. Computational politeness has long been a research strand for linguists \cite{danescu2013computational} as it leads to a better user experience with CUIs \cite{rana2021effect}. Whether by intentional design, a large corpus, or Reinforcement Learning from Human Feedback, ChatGPT seems to exhibit polite linguistic markers at scale and users recognize it as a positive. In addition to politeness, participants also commented on the substance of the output itself, which was much more relevant and focused than a search engine. However, it comes at the cost of having less variety in information that they can consume. In addition to the output being focused, participants recognized that ChatGPT can transform the output's complexity, so it is suitable for them. This ability to transform content complexity is perhaps the biggest pro for learning, in our opinion, as it aligns well with the instructional design goal of keeping a learner in the Zone of Proximal Development by adjusting (scaffolding) content so that it is not so easy that it is boring and not so hard that it is undigestible \cite{vygotsky1978mind, shabani2010vygotsky}. 

On the input side, while participants liked that the input to ChatGPT can be conversational, they also mentioned difficulties with finding the right input and noted that it is much easier to get information with Google just with a few keywords as opposed to prompt engineering an input for ChatGPT to get the desired output. However, participants also noted that it is beneficial to ask direct questions and receive direct responses with ChatGPT.

\subsection{Comfort, Confidence, Satisfaction, and Trust}

Satisfaction and comfort with interfaces are precursors to acceptability, usability, and user delight. Especially with AI, there have been issues with trusting the output due to biases \cite{memon2024search}, hallucinations \cite{rawte2023survey}, etc. rampantly exhibited by AI systems. For CUIs in particular, these errors would violate Grice's quality maxim \cite{grice1975logic} and lead to a hesitancy in tool acceptance \cite{choung2023trust}. In prior literature, educators have expressed that students might overlook the violation of maxims and instead believe the output of an AI without question \cite{Kiryakova2023}, while other researchers have found that users under-trust AI output and often to their own disadvantage \cite{walker2024they}. Our findings shed light on what causes comfort, satisfaction, confidence, trust, or distrust with information tools like ChatGPT and Google in the context of learning.

Participants mentioned that they feel more confident in their work in subsequent learning activities, such as class participation, after learning with ChatGPT. ChatGPT's focused output and ability to break down concepts into a customized level of reading with on-demand examples discussed in the previous sections potentially lead to a better understanding of the topic and consequential confidence in the material. However, participants mentioned some of that confidence is eroded because of the accuracy issues in ChatGPT's answers, lack of transparency in the process, and not always matching the user's output expectation. The lack of confidence is exacerbated when the user does not know enough about the domain they are learning and, as a result, does not feel confident about discerning true and false information. \citet{schneider2023investigating} mention that users with insufficient background knowledge find it difficult to express their goals to a conversational search agent. Google potentially solves this issue by providing control and transparency over sources of information.
Regarding Google, the participants mentioned using specific strategies to identify credible sources on Google. They can apply these strategies, as Google is a well-established learning tool whose pros and cons are discussed in many general education classes and digital literacy initiatives \cite{reddy2020digital,gutierrez2022digital}; especially at the institute where this research was conducted. As a result, students feel more confident in their learning; however, some explained that having to choose sources can hinder confidence, as it is another decision that one has to make. Confidence in Google is also eroded by ads masquerading as organic search results, a dark UX pattern that Google has previously been accused of in dark UX pattern registries \cite{unknown-author-2018}.

Regarding overall satisfaction, the most common theme of positive attitudes toward ChatGPT was its ability to synthesize information. Specifically, users liked that ChatGPT could prioritize information compared to its counterpart, which presents a wide set of concepts. Further, ChatGPT can connect the concepts, creating a single narrative. The users liked that they could understand one set of concepts presented by ChatGPT at a time without being flooded with seemingly disconnected pieces of information they could collect through Google search. For some, ChatGPT-generated information was easier than understanding explanatory diagrams available in search results.  Further participants stated that information could also be made easier to understand with a simple user request. However, when the prioritization of topics did not match the user's expectations, it led to lower satisfaction and abandonment. Then again, participants mentioned that the effort to look up information on ChatGPT was low enough that even if it didn't work as expected, it was worth checking. 

Overall comfort and satisfaction were also affected by the goal participants wanted to achieve. For example, participants mentioned that for well-established topics like the one in the study, they would use ChatGPT because there was not much room for misinformation, but for certain other topics or in-depth  reading, they would prefer Google so they can be exposed to various opinions instead of one narrative. \citet{sharma2024generative} have shown that LLMs can trap a user in an echo chamber. The fact that university students are aware of this limitation and actively avoid it is a positive result for the education community. Further, participants also mentioned that whether or not they use ChatGPT depends on the balance between wanting to thoroughly learn a topic and producing an output (essay), potentially acknowledging that the learning goal and the grade in university learning can diverge sometimes \cite{wiliam2001wrong} and ChatGPT can sometimes exclusively help with either achieving the goal or achieving the metric.

We also saw that while participants mentioned using ChatGPT, while straightforward,  also needed some specialized knowledge i.e., prompt engineering. Prompt engineering relates to an iterative process followed to provide inputs to a LLM such as ChatGPT in order to get relevant output from it \cite{reynolds2021prompt, giray2023prompt}. The participants mentioned that iterative application of prompting tips and tricks  helped to improve the experience. They mentioned that the uncertainty in knowing the right prompt and the time and effort required to engineer prompts was noticeable compared to Google where one could enter only a few keywords (rather than entire sentences and prompts) and receive a plethora of information. However,  we also heard that search results are not well synthesized and not adapted to user's reading level, and choosing sources in search is a pain point for users. 

It did not seem that privacy affected technology adoption for various reasons such as apathy, low-stakes data, benefit from giving up privacy, and feeling secure through shared vulnerability. However, data security still mattered to participants. Apathy and various attitudes towards privacy have been documented before \cite{hargittai2016can}, but now we also see it for AI applications. 

\subsection{How AI information spreads in the student community}
Related to tool adoption is how information about the tool and its use for academic activities spreads in the student community. Our participants noted that they first learned about AI through various sources, including co-workers, internships, friends, classmates, social media, news, professors, school teachers, and parents. When they first heard about AI, it was in various contexts, including, ironically, a message from a teacher not to use AI in the class. Most participants also learned about tips and tricks for using ChatGPT from the same or similar sources. 

\subsection{Learning with ChatGPT}
As discussed before, ChatGPT's ability to synthesize information and step up or down its complexity to keep a user in their zone of proximal development may help learning, as it is a well-established instructional design technique. In addition, participants mentioned ChatGPT helps provide structure to learning by creating a narrative and allowing a contextual follow-up to that narrative if a piece of information produced by ChatGPT is not clearly understandable. Such follow-ups can also include users asking for examples that illustrate the concept that ChatGPT can provide. Further, ChatGPT also provides writing support to the extent that students believe their writing has evolved due to using the tool. This is corroborated by research showing ChatGPT's ability to enhance the writing of students who speak English as a second language \cite{mahapatra2024impact}.

However, participants also mentioned complementary upsides of Google. For example, participants mentioned that they unintentionally discovered more topics via Google, whereas ChatGPT only gave a surface-level understanding. Further, when ChatGPT's output could not be trusted, participants mentioned using Google to find sources that could verify the information. 

Overall, participants thought the two tools were similar but not replaceable for learning. They discussed the pros and cons of each and mentioned using both tools to complete their learning tasks. This is also evident in a survey that showed that ChatGPT and Google are tools used almost equally in education and that one does not potentially replace the other \cite{divekar2024usage}. 

However, regarding a broader view of learning, the students mentioned that ChatGPT harmed their learning ability because it took away the mental exercise to find and synthesize disparate pieces of information. The participants mentioned that this leads to ``brain rust'' and ``laziness''. In addition, they mentioned that while they understood the topics at the moment, retention could be questionable because they did not put in as much hard work. The literature has shown that conversational learning interfaces can, in contrast, improve retention \cite{qiu2020conversational}. Further, we note that ease of learning does not directly mean that learning outcomes and retention will not be reached. In fact, being in the zone of proximal development can make the material easier to comprehend as opposed to reading at a level that the student cannot independently understand. However, some difficulties in learning are desirable and lead to growth \cite{bjork2006science}. Although some uses of ChatGPT, such as using it as a quiz partner, a personal tutor, or a brainstorming partner, could allow students to face \textit{desirable difficulties}, it is currently unclear whether general interactions with ChatGPT do this as \citet{bastani2024generative} have shown that specifically for math education, ChatGPT acted as a ``crutch'' that when taken away impacted students' skill. We note that the outcome may depend on the level and type of cognitive engagement with AI-generated information \cite{gajos2022people}.

\subsection{Academic Integrity and Plagiarism}
Notably, no final essay contained citations, even though they were written by university students who were in good academic standing and expected to know about citing their sources. We attribute this to the 10-minute time limit and experimental conditions in which the students wrote the final essay without expecting to submit it for a grade. Regardless of the situation, we asked the students their thoughts on academic integrity, plagiarism, and ownership in the context of ChatGPT and Google. 

One driver of not participating in plagiarism is the fear of getting caught. We noted that some students believed AI-generated content could be detected and, therefore, they could be caught in the act of plagiarism. However, research suggests there is no reliable way to detect AI-generated content either by humans \cite{casal2023can,fleckenstein2024teachers} or machines \cite{wu2023survey}. Other students were aware of that they cannot be caught using ChatGPT and thought that made ChatGPT as a more attractive option. Some others questioned whether ChatGPT's output was plagiarism at all. Students who wanted to avoid plagiarism by citing the source said that ChatGPT is difficult to use because the sources were not available. We note that products (e.g., Perplexity.ai, You.com, Microsoft Bing/Copilot, etc.)  incorporate LLMs with web search \cite{Wang2023LargeSM} and can cite  sources. They address the issue and have been available but not popular in the student community \cite{divekar2024usage}.  

Related to plagiarism is the concept of ownership over the produced material \cite{haviland2009owns}. We found various reasons why students felt they owned the final essay they produced, and it differed with the two tools. With Google, students felt ownership because they selected their sources, synthesized, and paraphrased the information. With ChatGPT, we found that students felt ownership when they verified the information. The verification step gave the students an agency over the final work produced. As one participant mentioned, ``If I was 100\% confident that it was always correct, I would feel almost no ownership.'' Some took the agency argument further and said that they owned the material generated by ChatGPT because it was their prompt that led to the generation. Others said that a human did not produce the content, so there was no one to attribute ownership except themselves. Yet others said they felt ownership when they synthesized the information from ChatGPT and wrote it in their own words in the final essay. 

The issue of plagiarism and ownership of AI-generated material is contentious. However, we see that the arguments for plagiarism and ownership are not in alignment with many university and educator policies where AI-generated content, when allowed, must be quoted and cited with attribution to the AI.

\section{Conclusion}

Given the rapid availability and usage of LLMs in education, this article sought to investigate the drivers, such as use cases, comfort, satisfaction, and trust, that lead to the adoption and use of these novel tools by students for general learning in higher education. We sought to compare student attitudes to these novel tools in light of one predominant technology in the student community, i.e., search engines. We pitted the two technologies against each other in a counterbalanced study to provide a context for a follow-up interview. Based on the qualitative analysis conducted on participant responses, we found that students appreciate ChatGPT's conversational interface, the ability to synthesize information, and the capacity to adjust content complexity to their level of understanding.  These features align with established instructional design principles and can potentially improve the learning experience. However, concerns about accuracy, lack of transparency in sources, and potential negative impacts on critical thinking and information synthesis skills persist. Google, while seen as more reliable due to its variety of sources and established credibility, is perceived as overwhelming in its information presentation. Students have developed strategies for source selection with Google, a skill that may be less applicable when using ChatGPT. Privacy concerns, while present, do not appear to significantly impact tool adoption, reflecting broader trends in attitudes toward digital privacy among young adults.

Interestingly, students often view these tools as complementary rather than mutually exclusive, using them strategically to consume and verify information and gain different perspectives. This suggests that instead of replacing traditional search engines, AI tools such as ChatGPT are being integrated into existing learning processes.

The study also sheds light on how information about AI tools spreads within student communities, highlighting the role of peers, social media, and educators in shaping adoption and use patterns. Furthermore, it reveals complex attitudes toward academic integrity and ownership of AI-generated content, which may not always align with institutional policies. 

As a result of these findings, educators can empathize with their students and will know how their students are being introduced to ChatGPT, how they use it for learning, how the learning process has changed from when students only had access to search engines, the depth of knowledge regarding these tools that students possess, and students' perspectives on academic integrity and ownership. This information will help educators design their learning materials, classroom policies, and learning expectations to address the new wave of the technology revolution in learning.


\bibliographystyle{ACM-Reference-Format}
\bibliography{sample-base}


\end{document}